\UseRawInputEncoding
\documentclass[10pt, conference, compsocconf]{IEEEtran}

\usepackage{float}
\usepackage{booktabs}
\usepackage{caption}
\usepackage{multirow}
\usepackage{graphicx}
\usepackage{caption}
\usepackage{subfig}
\usepackage{overpic}
\ifCLASSINFOpdf

\else
  
\fi

\begin{document}

\title{Exploring Turkish Speech Recognition via Hybrid CTC/Attention Architecture and Multi-feature Fusion Network}


\author{
    \IEEEauthorblockN{
        Zeyu Ren\IEEEauthorrefmark{2},
        Nurmement Yolwas\IEEEauthorrefmark{1},
        Huiru Wang\IEEEauthorrefmark{3},
        Wushour Slamu\IEEEauthorrefmark{2},
        \IEEEauthorblockA{
            \IEEEauthorrefmark{2}Xinjiang Multilingual Information Technology Laboratory, College of Information Science and Engineering \\
            Xinjiang University, Urumqi, China\\
            Email: renzeyu@stu.xju.edu.cn; wushour@xju.edu.cn  
        }
        \IEEEauthorblockA{
            \IEEEauthorrefmark{3}Xinjiang Key Laboratory of Signal Detection, College of Information Science and Engineering \\
            Xinjiang University, Urumqi, China\\
            Email: w176150@stu.xju.edu.cn
        }
        \IEEEauthorblockA{
            \IEEEauthorrefmark{1}Xinjiang Multilingual Information Technology 
            Research Center, College of Information Science and Engineering \\
            Xinjiang University, Urumqi, China\\ 
            Email: nurmemet@xju.edu.cn
        }
    }
}


%


\maketitle

\begin{abstract}
In recent years, End-to-End speech recognition technology based on deep learning has developed rapidly. Due to the lack of Turkish speech data, the performance of Turkish speech recognition system is poor. Firstly, this paper studies a series of speech recognition tuning technologies. The results show that the performance of the model is the best when the data enhancement technology combining speed perturbation with noise addition is adopted and the beam search width is set to 16. Secondly, to maximize the use of effective feature information and improve the accuracy of feature extraction, this paper proposes a new feature extractor LSPC. LSPC and LiGRU network are combined to form a shared encoder structure, and model compression is realized. The results show that the performance of LSPC is better than MSPC and VGGnet when only using Fbank features, and the WER is improved by 1.01\% and 2.53\% respectively. Finally, based on the above two points, a new multi-feature fusion network is proposed as the main structure of the encoder. The results show that the WER of the proposed feature fusion network based on LSPC is improved by 0.82\% and 1.94\% again compared with the single feature (Fbank feature and Spectrogram feature) extraction using LSPC. Our model achieves performance comparable to that of advanced End-to-End models.
\end{abstract}

\begin{IEEEkeywords}
End-to-End; Automatic speech recognition; Feature extractor; Feature fusion; LSPC
\end{IEEEkeywords}

%
\IEEEpeerreviewmaketitle

\section{Introduction}
Speech recognition is an important research area in speech signal processing, with a wide range of research significance and applications. Automatic Speech Recognition (ASR), also known as speech recognition, converts a speech signal containing certain speech information into a text sequence using algorithms implemented by a computer program. Speech recognition technology is used in a wide range of applications, for example in the smart home sector, where users can control home devices with voice commands, thus enabling intelligent home living. In the driverless space, speech recognition technology can help vehicles recognize the commands of the driver, thus enabling a smarter driving experience. In the field of chatbots, speech recognition technology can help robots understand the user's language and therefore provide better services to the user. Speech recognition technology has significant application value and development prospects in modern society. In the future, as technology advances and applications continue to expand, speech recognition technology will be used and developed in more and more areas.

Turkish mainly employs a fusion of affixes and roots in word formation. The same affix combines with different roots to produce different word meanings, and the same affix has different grammatical meanings in different sentence forms. Turkish has a rich vocabulary as a result of attaching various grammatical components to its roots as a means of word formation and morphology, which makes speech recognition more challenging with limited annotation data. In the field of deep learning, models complete their training by learning vast amounts of data and are unable to exclude factors such as noise in the same way as humans, so the lower the coverage in everyday scenarios, the lower the recognition accuracy, leading to a failure to land applications. The complexity and distortion of words lead to an increased word error rate (WER), making speech recognition difficult.

For agglutinative language, including Turkish, researchers have conducted a series of studies. In 2016, Dawel et al.\cite{Dawel2016on} constructed a GMM-HMM based continuous speech recognition system for Kazakh. Since 2019, Al-Farabi Kazakh National University has built a DNN-HMM system\cite{mamyrbayev2019continuous}, a BLSTM-CTC End-to-End system\cite{mamyrbayev2020end}, and a Transformer system\cite{orken2022study} for the Kazakh language using private data in succession. Meanwhile, Beibut et al.\cite{beibut2020development} constructed a transfer learning-based LSTM-CTC End-to-End ASR system for Kazakh. In 2020, Musaev et al.\cite{musaev2020use} completed modeling by combining convolutional neural networks and long and short-term memory networks and used multiple features to improve speech recognition accuracy in Uzbek. Çetinkaya et al.\cite{ccetinkaya2020improving} used a subword-based modeling approach with optimization techniques such as model regularisation to improve data sparsity and recognition errorr problems in Turkish speech recognition.

In summary, although End-to-End speech recognition technology is developing rapidly, it has been less applied in adherent languages and still faces many challenges. Therefore, how to apply End-to-End speech recognition technology to adjective languages, how to make use of the scalability and flexibility of End-to-End speech recognized technology to further improve the performance of models and speech recognition accuracy in these languages, and how to solve the difficulties of traditional speech recognition technology in handling adjective languages are the key issues to be studied.

In this paper, the training efficiency and accuracy of the model for speech recognition are investigated based on a hybrid CTC/Attention structure. The main contributions of this paper are as follows:

• Based on previous work in \cite{ren2022improving}, the effects of different data enhancement methods and decoding beam search widths on model performance are explored to find the best optimization strategy to improve the generalization of the model and to obtain higher accuracy as well as optimization of the model.
• A new feature extractor, LSPC, is proposed to capture the features of the input data at a more detailed level, allowing the model to better learn the features of the data, thus improving the generalization capability of the model.

• We construct a multi-feature speech recognition model, propose a new feature fusion network to maximize mise the use of adequate feature information, and combine the feature fusion network with LiGRU to form the final shared encoder structure.


\section{Related Work}
Traditional methods require separate training of modules such as acoustic models, pronunciation dictionaries, and language models, which is very complex and requires support from alignment models. On the other hand, End-to-End automatic speech recognition is a single integrated approach that uses a data-driven approach, thereby reducing the complexity of the overall speech recognition system. Specifically, End-to-End automatic speech recognition treats the speech recognition problem as a sequence-to-sequence problem, by taking the speech signal as the input sequence and transcribing the text as the output sequence, and directly delegating the entire speech recognition task to a deep neural network. This approach automatically learns the correspondence between audio and text using large amounts of training data and allows text to be inferred directly from the speech signal without the need for a cumbersome alignment model.

End-to-End approach mainly includes Attention-based Encoder–Decoder model\cite{zeyer2018improved} (AED), Connectionist Temporal Classification model\cite{graves2006connectionist} (CTC) and Recurrent Neural Network Transducer model\cite{li2019improving} (RNN-T). CTC uses the addition of a blank tag \{blank\} to the output sequence to align the speech frame sequence with the text sequence to simplify the training process. Unlike the HMM structure, it automatically learns and optimizes the correspondence between audio information and annotated text during training and does not require frame alignment to be achieved before the network is trained. The disadvantage is that it assumes that each token is independent of the others, but in reality, there are contextual dependencies between the tokens. To address this problem, RNN-T introduces a predictive network to learn contextual information, which is equivalent to a language model. Compared to CTC, RNN-T is more difficult to train, although there is no longer a restriction on the length of the input and output sequences. Another way to mitigate the conditional independence assumption is to use an attention mechanism\cite{das2018advancing}, which allows the model to better focus on the acoustic features associated with the target text, improving the encoder structure to mitigate the conditional independence assumption without changing the objective function. AED, another of the most commonly used architectures, contains an encoder module for extracting features and a decoder module that uses an attention mechanism. This architecture can be used with various types of neural networks. In contrast to CTC, AED does not require conditional independence assumptions and uses a recurrent neural network and an attention module to construct the encoder and decoder to achieve soft alignment between labels and audio information. However, the AED model is less able to generalize to long audio segments. The inconsistent length of the input and output sequences increases the difficulty of alignment. For long audio, a window must be manually set to limit the scope of attentional exploration. Moreover, alignment in the attention mechanism is easily corrupted by noise.

In order to solve the alignment problem in AED, Kim S et al. proposed a joint CTC/Attention structure\cite{kim2017joint}, which uses a joint training approach and considers both CTC and attention mechanisms in the training process. The model uses both CTC and attention mechanisms, and can effectively address errors caused by CTC being too conservative or attention mechanisms being too free. It accelerates the convergence of the model while correcting alignment problems and has become the standard training approach for most AED models\cite{liu2019adversarial}. In the \cite{hori2017joint}, further improvements were made to the model by combining scores from both AED and CTC models in both rescoring and one-pass decoding during decoding. Seki H et al. accelerated the hybrid by vectorising multiple hypotheses in beam search decoding process of the CTC/Attention model\cite{seki2019vectorized}. Subsequently, various hybrid models were proposed to solve the alignment problem\cite{moritz2019triggered}.

Although End-to-End speech recognition techniques are developing rapidly, they are less commonly used in Turkish and still face many challenges. Therefore, the application of End-to-End speech recognition technology to Turkish, the scalability and flexibility of the technology to further improve the performance of models and speech recognition accuracy have become key issues for research.
\section{Model}
\begin{figure}[!htbp]
\includegraphics[width=0.5\textwidth]{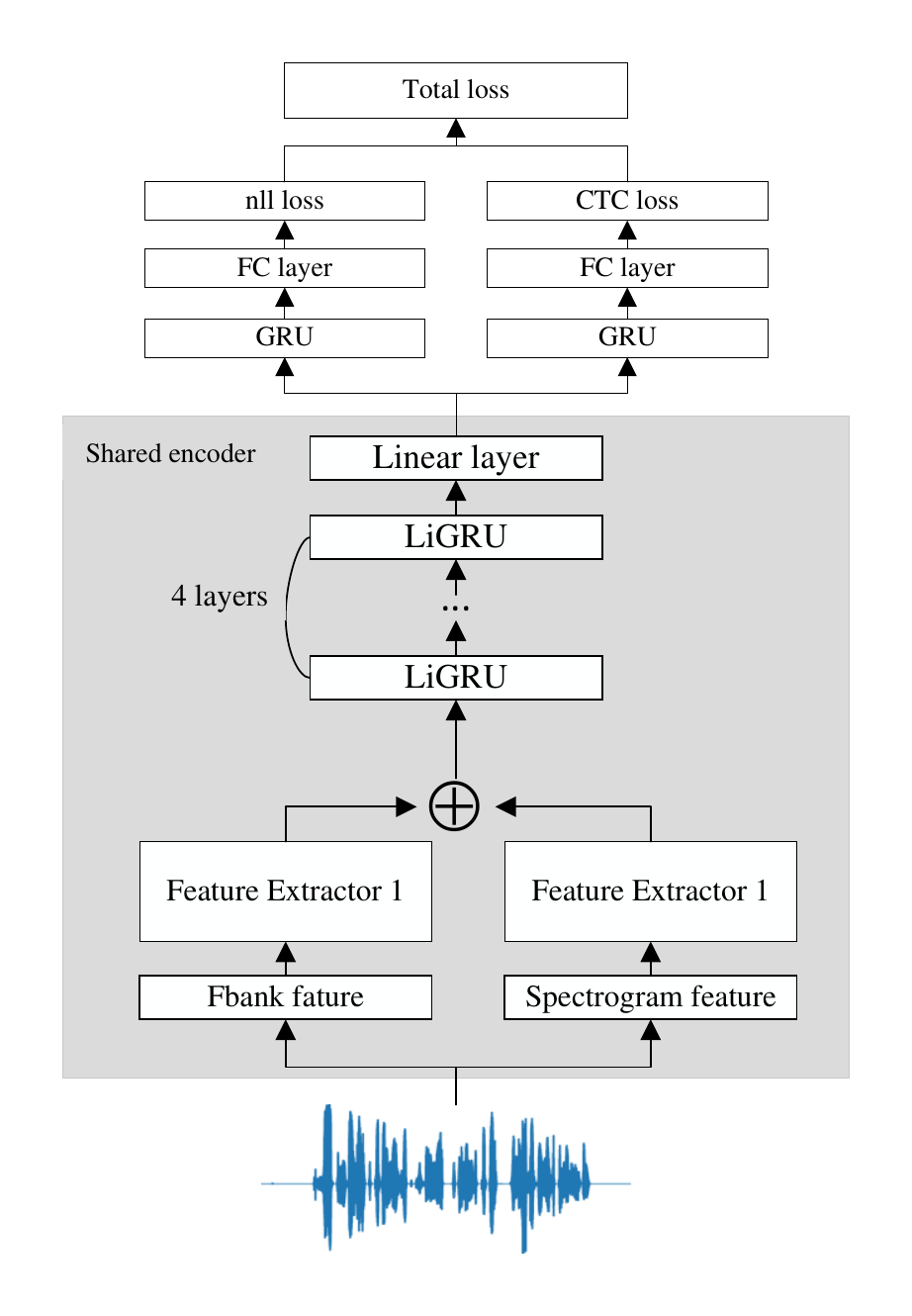}
\caption{Hybrid CTC/Attention Model Structure Based on Feature Fusion}
\label{fig1}
\end{figure}

The hybrid CTC/Attention model generally consists of two components: shared encoder module and joint decoder module. The shared encoder module converts the speech signal into a sequence of feature vectors, typically using a model such as a Convolutional Neural Network (CNN) or Recurrent Neural Network (RNN). Joint decoder consisting of a CTC decoder and an Attention decoder. The CTC module performs character-level alignment of the feature vector sequence, producing a series of characters as an intermediate result, consisting of multiple fully-connected layers. The final layer outputs a character sequence that represents the character-level alignment result of the input speech signal. The Attention module is generally consists of multiple RNN or Transformer layers, which is used to achieve word-level alignment and produce the final recognition results. The specific model structure is shown in Figure. \ref{fig1}. Where the nll loss denotes the negative log-likelihood loss.

\subsection{Shared encoder}
The encoder network consists mainly of convolutional neural network (CNN) based feature extractor and a set of bi-directional LiGRU networks.
\subsubsection{Feature extractor}
\begin{figure}[!htbp]
\includegraphics[width=0.5\textwidth]{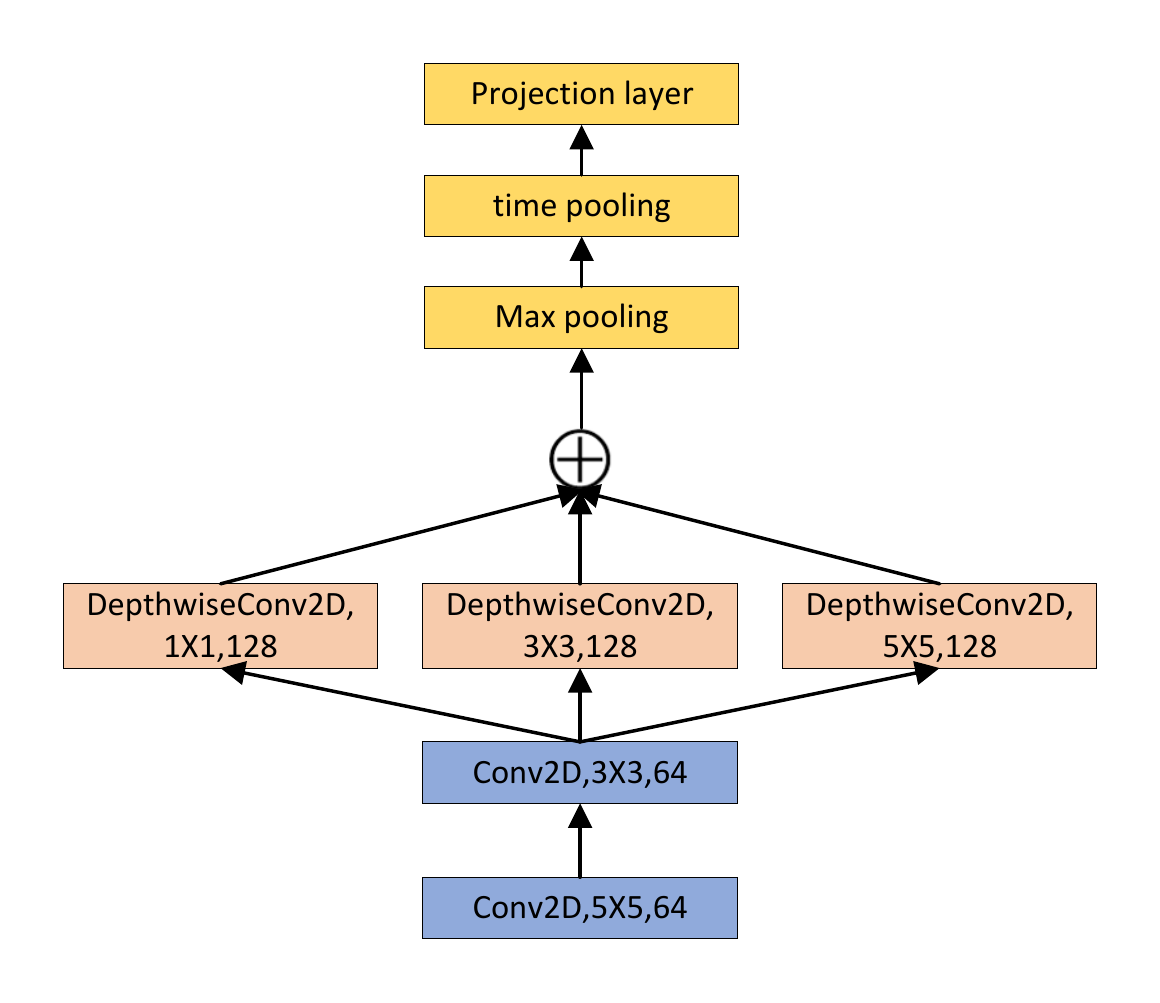}
\caption{Detailed architecture of the feature extractor LSPC}
\label{fig2}
\end{figure}

\ 
\newline
\indent We propose a lightweight multi-scale parallel convolution (LSPC) based on deep CNN networks, which consists of 2 CNN layers, 2 pooling layers, 1 set of parallel CNN layers, and 1 fully connected layer. The detailed structure of the LSPC is shown in Figure. \ref{fig2}. 

As the datasets used are transcribed data collected from different speakers, there will be differences in speech speed, pronunciation, intonation, etc. To capture features at different time scales, parallel convolution is incorporated in the design of the feature extractor network, which extracts features at multiple scales by processing several CNN layers with different kernel sizes in parallel. 

After parallel convolution, the features extracted from the three branches are stitched together to form the final feature representation. Next, two consecutive 1D pooling layers are connected to perform pooling operations on the second-dimensional data and the temporal dimension respectively. This is designed with the following reasons in mind: firstly, the 1D pooling layer only needs to perform operations in one direction, which is faster to compute and reduces the number of parameters of the model compared to the 2D pooling layer. Secondly, the use of two 1D pooling layers allows the features of the input data to be captured at a more detailed level, allowing the model to better learn the features of the data, thus improving the generalization capability of the model. Finally, this design allows different pooling sizes to be set according to the different dimensions of the input data to better accommodate different input data.

After the pooling layers, we connected a linear projection layer with an output unit of 1024 to control the dimensionality of the output of the convolution and pooling layers, thus effectively controlling the number of parameters. The non-linear capability of the LSPC is increased by mapping the output features to a suitable space, thus improving the expressiveness and extraction accuracy of the feature extractor. 
\subsubsection{Feature fusion network}
\ 
\newline
\indent Fbank\cite{pardede2019generalized} feature and Spectrogram\cite{meghanani2021exploration} feature are input into a separate feature extractor LSPC, which can not only fully extract different feature vectors, but also avoid mutual interference between different features. After weighted fusion, the output vectors are input into the 4 layer LiGRU network and then connected with two fully connection layers to reduce the dimension of the time series features of LiGRU output, to obtain a more compact and differentiated feature representation.

See Equations (\ref{eq:1})$\sim$(\ref{eq:4}) for the specific implementation process of the shared encoder.
\begin{equation}
    f_1=LSPC(featu{re}_F) \label{eq:1}
\end{equation}
\begin{equation}
   f_2=LSPC(featu{re}_s) \label{eq:2}
\end{equation}
\begin{equation}
   f=\left(1-\beta\right)\ast f_1+\beta\ast f_2 \label{eq:3}
\end{equation}
\begin{equation}
   h_t^{enc}=f_{enc}\left(X\right)=Lin\left({LiGRU}^{4\times}\left(f\right)\right) \label{eq:4}
\end{equation}

$featu {re} _ F$ and $featu {re} _ s$ represent Fbank feature and Spectrogram feature extracted from the original speech $X$, and $f_1$ and $f_2$ are obtained by feature extractor LSPC. 
Vectors $f_1$ and $f_2$ are fused into vector $f$ by parameter $\beta$. This parameter is automatically adjusted during training by the torch.nn.Parameter(·) function, with an initial value of 0.5. Through the encoder structure, the high-level feature representation $h_t^{enc}$ is finally obtained. 
\subsection{Joint decoder}
The joint decoder consists mainly of CTC decoder and Attention decoder. The CTC loss function is calculated as follows Eq.(\ref{eq:5})$\sim$(\ref{eq:7}):
\begin{equation}
    loss_{ctc}=-lnP\left(\left.y\right|x\right) \label{eq:5}
\end{equation}
\begin{equation}
    P\left(\left.y\right|x\right)=\sum_{\pi\in Q\left(y\right)}P\left(\left.\pi\right|x\right) \label{eq:6}
\end{equation}
\begin{equation}
    P\left(\left.\pi\right|x\right)=\prod_{t=1}^{T}P\left(\left.q_t\left(y_t\right)\right|x\right) \label{eq:7}
\end{equation}

where, $\pi$ is one of the CTC paths. The following Eq. (\ref{eq:8}) shows the loss function of the attention module:
\begin{equation}
 loss_{Att}=-lnP\left(\left.y^\ast\right|x\right)=-\sum_{u}{ln P\left(\left.y_u^\ast\right|x,y_{1:u-1}^\ast\right)} \label{eq:8}
\end{equation}

where $y_{1:u-1}^\ast$ denotes the true label of the previous character.The linear combination between the CTC loss function in Eq. (\ref{eq:5}) and the Attention loss function in Eq. (\ref{eq:8}) using the parameter $\lambda$: 
\begin{equation}
 loss_{hybrid}=\lambda loss_{CTC}+\left(1-\lambda\right)loss_{Att} \label{eq:9}
\end{equation}

The decoder uses joint decoding method, including two branches. CTC participates in auxiliary decoding, which helps the speech signal to be properly aligned in noisy environment, improves the unstable training of attention under noise interference, and improves the convergence speed of the model.

The objective function of the joint decoding can be expressed as Eq. (\ref{eq:10}):
\begin{equation}
 \hat{c}=\arg \max _{c \in u^\ast}\left\{\lambda \log P_{CTC}(c \mid x)+(1-\lambda) \log P_{Att}(c \mid x)\right\} \label{eq:10}
\end{equation}

Where $0\le\lambda\le1$ and we let $\lambda=0.3$. Because the Attention decoder is performed with the output tags synchronized, whereas the CTC is performed in frame synchronization mode. To include the probability of CTC in the score, we use a one-time decoding algorithm. The algorithm combines CTC and Attention modules, uses the scores of candidate output sequences to calculate the output results, and can balance the influence of the two by adjusting the super parameter $\lambda$. The beam search method is used to quickly find the sequence with the highest score as output without enumerating all possible output sequences. The CTC and Attention modules are used to calculate the probability of their respective inferences, and the final result is obtained by linearly combining the CTC-based score $\alpha_ {CTC}\left(h, x\right)$ and the Attention-based score $\alpha_{Att}\left(h,x\right)$ through the parameter $\lambda$, as shown in formula (\ref{eq:11}):
\begin{equation}
 \hat{c}=\arg \max _{h \in \hat{P}}\left\{\lambda \alpha_{CTC}(h,x)+(1-\lambda) \alpha_{Att}(h,x)\right\} \label{eq:11}
\end{equation}

\section{Experiments}
In this section, we first introduce the Turkish speech recognition dataset, then describe the relevant models in detail and carry out the related experiments. In the experiments, we first explore new model optimization strategies to achieve the best results. Finally, the feature extractor LSPC and multi-feature fusion network proposed in this paper are tested and analyzed.
\subsection{data preparation}
The Common Voice speech dataset\cite{ardila2020common} is a multilingual public dataset containing 17,127 hours of speech data in 104 languages, with each entry in the dataset consisting of a unique MP3 audio file and a corresponding text file. The Turkish corpus used in our study was collected and validated through Mozilla's Common Voice initiative.The specific information of the data set is shown in Table \ref{TableA}.
\begin{table}[!htbp]
\centering
\caption{\textbf{Details of dataset corpus allocation}}
\label{TableA}
\begin{tabular}{cccccc}
    \toprule
    \multirow{2.5}{*}{datatset} & \multicolumn{3}{c}{duration(h)} &  \multirow{2.5}{*}{total duration(h)} & \multirow{2.5}{*}{total speakers} \\
    \cmidrule{2-4}
    & train & dev & test\\ 
    \midrule
    cv8-tr & 16.36 & 8.53 & 9.65 & 34.54 & 1264 \\
\bottomrule
\end{tabular}
\end{table}
 
 80-dimensional Fbank feature were extracted for 25ms long and 10ms shifted speech frames. Spectrogram feature used 201-dimensional.
\subsection{Exploring optimal model optimisation strategies}
In this section, a series of optimization strategies are explored for the improved model of the hybrid CTC/Attention structure in reference\cite{ren2022improving} to achieve the best performance of the model.
\subsubsection{Exploration of the best data enhancement option}
\ 
\newline
\indent The Turkish corpus as a low resource corpus often does not cover all speech and environmental variation and collecting and annotating large amounts of speech data is difficult and expensive. By using data augmentation, it is possible to increase the amount and diversity of speech data and reduce the risk of overfitting training without increasing the cost of data collection and annotation. Table \ref{TableB} shows comparison of the experimental results for different combinations of SpecAugment\cite{park2019specaugment}, Speed perturbation, and noise addition\cite{snyder2015musan} data enhancement methods for beam width of 8.

\begin{table}[!htbp]
    \centering
    \caption{\textbf{Results of using different data augmentation combinations}}
    \label{TableB}
    \begin{tabular}{ccc}
        \toprule
        Method & CER (\%) & WER (\%) \\
        \midrule
        SpecAugment & 24.30 & 53.35 \\ 
        Speed perturbation & 23.46 & 56.11 \\ 
        Noisy & 29.14 & 60.12 \\ 
        SpecAugment + Speed perturbation & 29.75 & 60.11 \\ 
        SpecAugment + Noisy & 23.30 & 51.44 \\ 
        Speed perturbation + Noisy & 24.39 & 50.91 \\ 
        \bottomrule
    \end{tabular}
\end{table}
As can be seen from Table \ref{TableB}, when using only the SpecAugment data enhancement method, it achieves good results compared with the other two methods, WER is reduced by 2.76\% and 6.77\%.

When the SpecAugment and speed perturbation are used together, the experimental results are not ideal. This may be because both Specaugment and Speed perturbation methods transform the speech signal in the time domain, and if they are used at the same time, it may lead to repeated enhancement of data. However, when the noise adding method is used in combination with SpecAugment or Speed perturbation, the data enhancement effect is very good. This is because the data enhancement methods are complementary. The method of adding noise is to transform in the amplitude domain of speech signal, which can simulate some actual noise and interference, while the method of Specaugment and Speed perturbation is to transform in the frequency domain or time domain, which can simulate different changes in some speech signals.

In contrast, the combination of Speed perturbation and noise adding enhancement improves the WER by 2.44\% compared with the SpecAugment method alone. Compared with other methods, the advantages are also obvious.

\subsubsection{Exploration of optimal beam search width}
\ 
\newline
\indent After finding the best data enhancement strategy, we explore different beam widths to achieve a balance between decoding accuracy and decoding efficiency. This section compares and analyzes the experimental results when the beam search width is 8, 12 and 16. See Table \ref{TableC} for details.

\begin{table*}[!htbp]
\centering
\caption{\textbf{Comparison of experimental results under different beam widths}}
\label{TableC}
\tabcolsep=0.016\linewidth
\begin{tabular}{ccccccccc}

    \toprule
    \multirow{2.5}{*}{Method(no LM)} & \multicolumn{2}{c}{width=8}  & \multicolumn{2}{c}{width=12} & \multicolumn{2}{c}{width=16}\\
    \cmidrule{2-7}
    & CER(\%) & WER(\%) & CER(\%) & WER(\%) & CER(\%) & WER(\%)\\
    \midrule
    Specaugment & 24.30 & 53.35 & 23.43 & 53.41 & 23.42 & 52.77 \\
    Speed perturbation & 23.46 & 56.11 & 23.15 & 55.54 & 23.02 & 54.76 \\ 
    Specaugment+noisy & 23.30 & 51.44 & 22.56 & 50.64 & 23.17 & 50.81 \\ 
    Speed perturbation+noisy & 24.39 & 50.91 & 19.18 & 47.70 & 19.18 & 47.59 \\ 
    Speed perturbation+noisy+LM & 19.23 & 48.01 & 18.58 & 47.00 & 18.47 & 46.78 \\ 
\bottomrule
\end{tabular}
\end{table*}

\begin{figure}[htbp]
	\centering
	\subfloat[Comparison of data enhancement methods WERs]{\includegraphics[width=1.0\columnwidth]{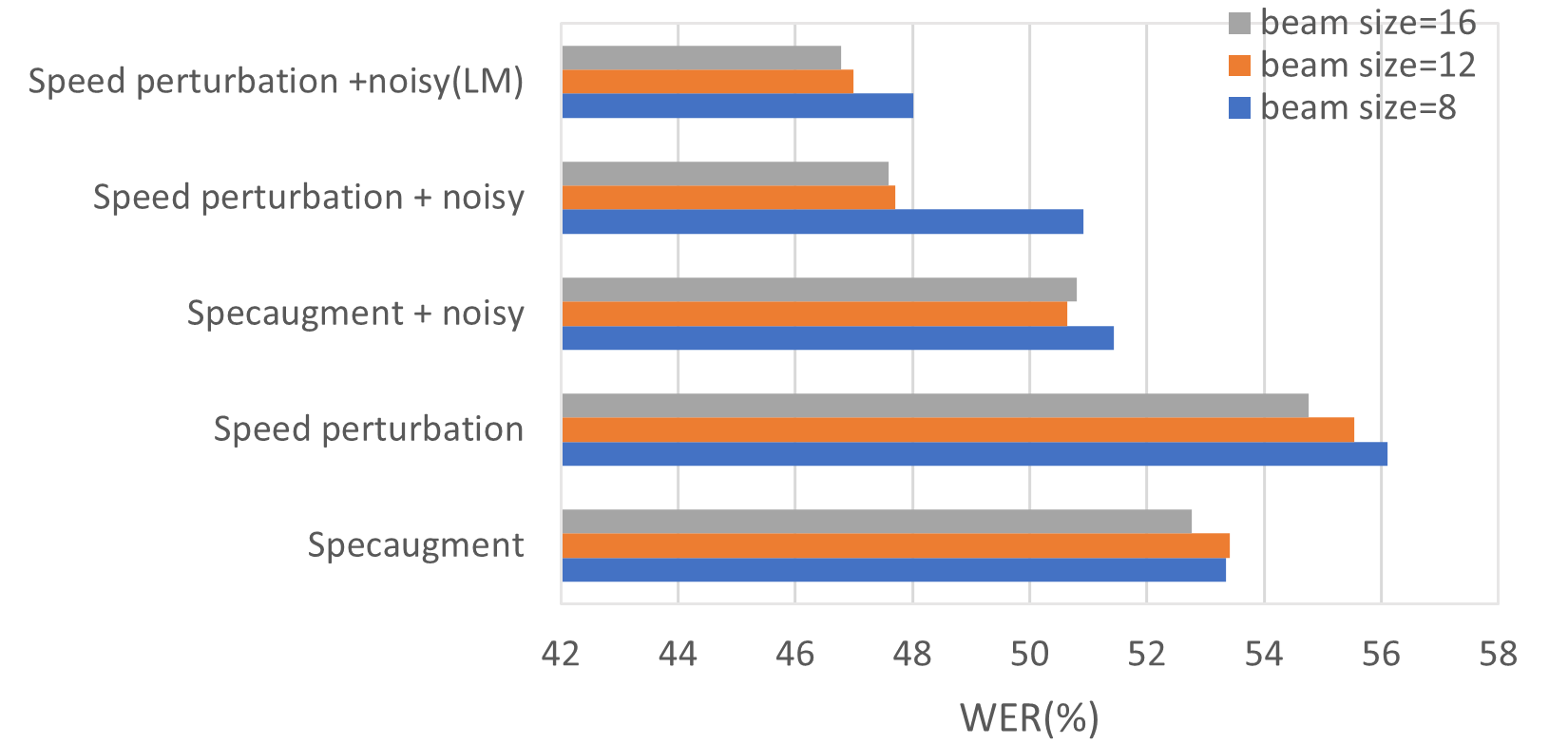}}\hspace{5pt}
	\\
	\subfloat[Comparison of data enhancement methods CERs]{\includegraphics[width=1.0\columnwidth]{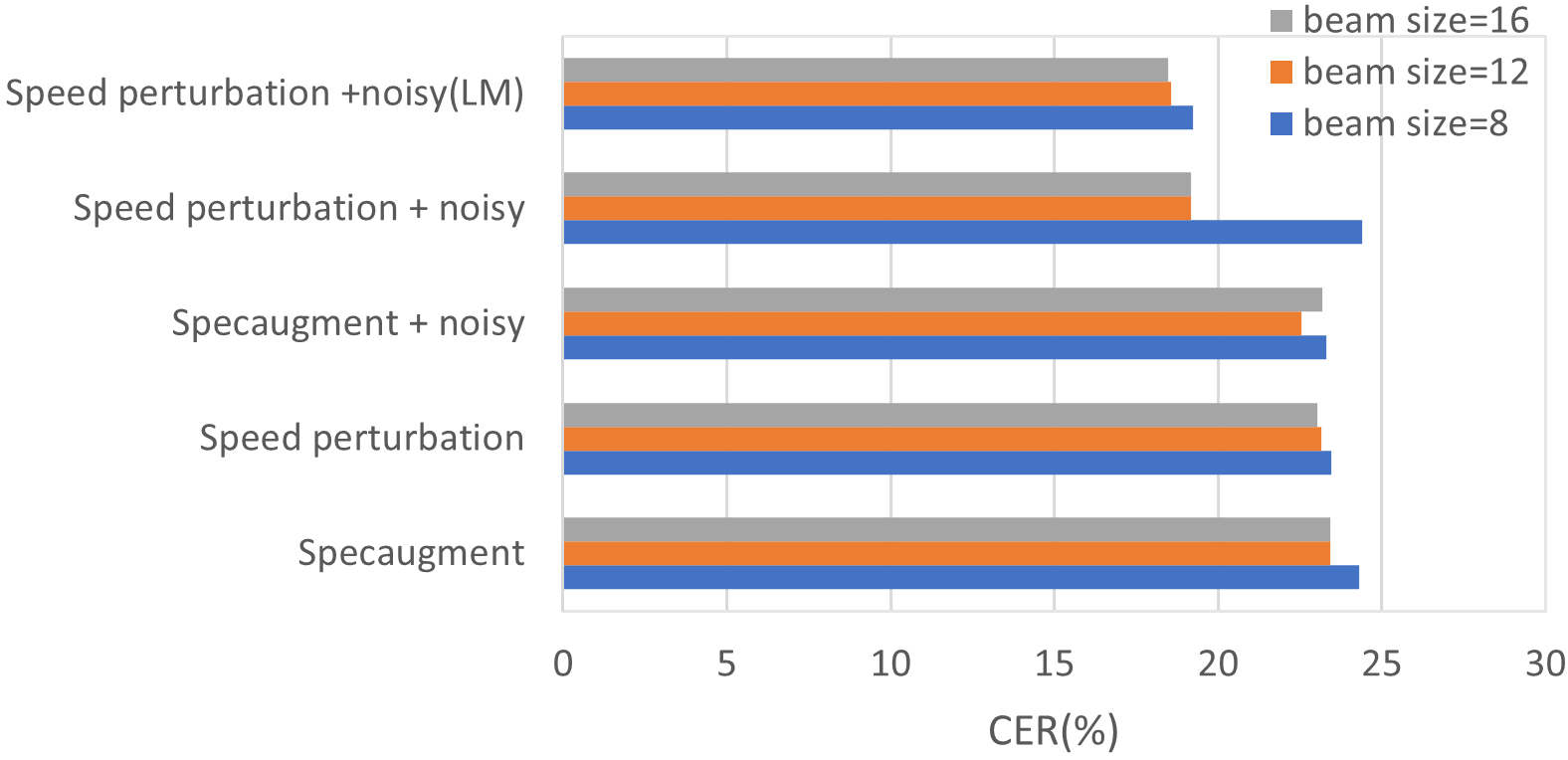}}
	\caption{Comparison of data enhancement effects under different beam widths}
        \label{fig:3}
\end{figure}

 The results in Table \ref{TableC} and Figure. \ref{fig:3} show that CER and WER decreases progressively with increasing set beam width during decoding. When larger widths are used, there is a definite improvement in system performance. This is because, when the beamwidth is 8, the search space is more limited and the system may miss some possible recognition paths. Whereas, when the beam width is 16, the search space is wider and the system can explore different recognition paths more fully, thus finding better recognition results.

With the inclusion of the language model, CER and WER was reduced by approximately 5\% and 3\% respectively when the beam width was 8. The relative improvement was further reduced by the set width when the beam width was increased to 16. This is because problems such as ambiguity or noise in the recognition of speech signals make it difficult for the decoder to find a globally optimal solution. However, when a pre-trained initialized language model is added, it can help the decoder to better understand the contextual information, constrain the search space for decoding and reduce decoding error. Therefore, under the combined consideration of decoding speed and accuracy, 16 is finally chosen as the optimal width.

\subsection{Effectiveness of the LSPC feature extractor}
In this section, ablation experiments were carried out to prove the effectiveness of the proposed LSPC structure and compared with MSPC. Then, the ability of LSPC and other extractors to extract depth features from Fbank feature and Spectrogram feature is verified.
\subsubsection{Ablation experiments}
\ 
\newline
\indent Table \ref{TableD} shows the ablation experimental structure of LSPC. LSPC + BLSTM + GRU means that LSPC and BLSTM are combined as a shared encoder and GRU as a decoder. All the experiments in Table 4 only experiment with the SpecAugment data enhancement method.
\begin{table}[!ht]
    \centering
    \caption{\textbf{Results of the ablation study of the proposed LSPC architecture}}
    \label{TableD}
    \begin{tabular}{ccc}
    
    \toprule
        Methods & CER(\%) & WER(\%) \\ 
        \midrule
        MSPC+BLSTM+GRU+location-based attention & 25.39 & 53.83 \\ 
        MSPC+BLSTM+GRU +content-based attention & 24.35 & 54.61 \\ 
        MSPC+BLSTM+GRU+location-LSTM attention & 24.30 & 53.35 \\ 
        \midrule
        LSPC+BLSTM+GRU+location-based attention & 21.07 & 49.46 \\ 
        LSPC+BLSTM+GRU+content-based attention & 21.50 & 50.23 \\ 
        LSPC+BLSTM+GRU+location-LSTM attention & 21.18 & 49.16 \\ 
    \bottomrule
    \end{tabular}
\end{table}

As you can see from Table \ref{TableD}, LSPC works best when used as a feature extractor in conjunction with Location-LSTM attention.
\subsubsection{Comparison of different feature extractors }
\ 
\newline
\indent In this section, the proposed new feature extractor LSPC is experimented with to extract deep features from Fbank and Spectrogram features, which verifies the effectiveness and performance improvement of this structure compared with feature extractors VGGnet and MSPC. The results are shown in Table \ref{TableE}.
\begin{table}[!ht]
    \centering
    \caption{\textbf{Comparison of different feature extractors}}
    \label{TableE}
    \begin{tabular}{cccccc}
    \toprule
        ~ & Feature & Extractor & CER (\%) & WER (\%) & Parameters \\ 
        \midrule
        1 & Fbank & VGGnet & 19.83 & 50.11 & 144.8M  \\ 
        2 & Fbank & MSPC & 19.60 & 48.59 & 224.2M  \\ 
        3 & Fbank & LSPC & 18.45 & 47.58 & 57.2M  \\ 
        4 & Spectrogram & VGGnet & 20.49 & 50.30 & 155.0M  \\ 
        5 & Spectrogram & MSPC & 23.06 & 51.71 & 418.0M  \\ 
        6 & Spectrogram & LSPC & 19.58 & 48.70 & 81.0M  \\ 
    \bottomrule
    \end{tabular}
\end{table}

As can be seen from Table \ref{TableE}, by comparing the experimental results, it is found that when using Fbank features, the feature extractor proposed in this chapter is improved by 1.38\% and 2.53\% compared with VGGnet, CER and WER, respectively, and the parameters are reduced by 87.6M. Compared with MSPC, CER and WER increased by 1.15\% and 1.01\% respectively, and the parameters decreased by 167M. When using Spectrogram feature, WER is reduced by 1.6\% and 2.99\% compared with VGGnet and MSPC, respectively. These results show the superiority of LSPC in feature extraction, which can better extract the information of speech data and improve the model performance. At the same time, LSPC also has obvious advantages in parameters, which can realize model compression and reduce model storage and calculation overhead.
\subsection{Effectiveness of Feature Fusion Network}
In this section, each feature extractor is used in the encoder network based on feature fusion proposed in this chapter to verify the effectiveness of our proposed fusion network. The experiments in this section use a data enhancement method combining Speed perturbation and noise addition. Table \ref{TableF} shows the results of each experiment.
\begin{table*}[!ht]
    \centering
    \caption{\textbf{Comparison of experimental results under different beam widths}}
    \label{TableF}
    \tabcolsep=0.016\linewidth
    \begin{tabular}{ccccccccc}
    \toprule
        ~ & Phonetic features & Beta & Extractor & With LM & CER(\%) & WER(\%) & Parameters \\ \midrule
        1 & Fbank+Spectrogram & 0.3 & VGGnet & no & 21.88 & 53.01 & 64.5M \\ 
        2 & Fbank+Spectrogram & 0.3 & LSPC & no & 18.56 & 46.90 & 97.8M \\ 
        3 & Fbank+Spectrogram & automatic & LSPC & no & 18.23 & 47.30 & 97.8M \\ 
        4 & Fbank+Spectrogram & automatic & LSPC & yes & 17.94 & 46.76 & 151.4M \\ 
    \bottomrule
    \end{tabular}
\end{table*}
Under the condition that LSPC is better than MSPC, the Fbank + Spectrogram feature is used in the feature fusion network, and the performance of feature extraction using VGGnet is compared. Combined with the results in Table \ref{TableE}, it can be seen that when using LSPC to extract deep features of Fbank feature and Spectrogram feature, the CER and WER of the former are 1.13\% and 1.12\% higher than those of the latter, respectively. Therefore, we still use Fbank features mainly. First, we use $\beta=0.3$ for feature fusion, and then we use a torch.nn.Parameter(·) function to automatically adjust during training, with an initial value of 0.5.

As can be seen from Table \ref{TableF}, when LSPC is used for feature extraction and two fusion parameters are used for fusion, the character error rate (CER) and word error rate (WER) are absolutely improved compared with VGGnet. In Experiment 4, the performance of the model is further improved after the language model is fused in decoding at a ratio of 0.5. This proves that LSPC can extract the most accurate information from the original multiple feature sets, and can improve the efficiency of real-time decision-making by removing redundant information between different feature sets, which proves the effectiveness of our proposed fusion network.
\section{Conclusion}
This paper studies a series of speech recognition tuning methods, including the data enhancement method and beam width. In addition, to improve the accuracy of feature extraction, a new LSPC feature extractor is proposed and related experiments are carried out. Finally, combines the LSPC feature extractor and shared encoder network based on feature fusion, and achieves absolute performance improvement. These research results are helpful to improve the accuracy and efficiency of speech recognition and are of great significance to research and practice in related fields.

We can also add the MFCC feature to the feature fusion network to fuse more deep semantic information and improve the robustness of the model. Besides feature selection, we can also consider building a multi-lingual corpus of agglutinative language to train the model, to achieve better results.

\section*{Acknowledgment}
This work was supported by the National Natural Science Foundation of China (Grant No.U1603262 and 62066043) and  the National Language Commission key Project(Grant No.ZDI135-133)

\bibliographystyle{IEEEtran}

\bibliography{RA}

\begin{thebibliography}{10}
\providecommand{\url}[1]{#1}
\csname url@samestyle\endcsname
\providecommand{\newblock}{\relax}
\providecommand{\bibinfo}[2]{#2}
\providecommand{\BIBentrySTDinterwordspacing}{\spaceskip=0pt\relax}
\providecommand{\BIBentryALTinterwordstretchfactor}{4}
\providecommand{\BIBentryALTinterwordspacing}{\spaceskip=\fontdimen2\font plus
\BIBentryALTinterwordstretchfactor\fontdimen3\font minus
  \fontdimen4\font\relax}
\providecommand{\BIBforeignlanguage}[2]{{%
\expandafter\ifx\csname l@#1\endcsname\relax
\typeout{** WARNING: IEEEtran.bst: No hyphenation pattern has been}%
\typeout{** loaded for the language `#1'. Using the pattern for}%
\typeout{** the default language instead.}%
\else
\language=\csname l@#1\endcsname
\fi
#2}}
\providecommand{\BIBdecl}{\relax}
\BIBdecl

\bibitem{Dawel2016on}
L.~Y. Dawel~Abilhayer, Nurmemet~Yolwas, ``On language model construction for
  lvcsr in kazakh,'' in \emph{Computer Engineering and Applications}, vol.
  52(24), 2016, pp. 178--181.

\bibitem{mamyrbayev2019continuous}
Ð.~Mamyrbayev, M.~Turdalyuly, N.~Mekebayev, K.~Mukhsina, A.~Keylan, B.~BabaAli,
  G.~Nabieva, A.~Duisenbayeva, and B.~Akhmetov, ``Continuous speech recognition
  of kazakh language,'' in \emph{ITM Web of Conferences}, vol.~24.\hskip 1em
  plus 0.5em minus 0.4em\relax EDP Sciences, 2019, p. 01012.

\bibitem{mamyrbayev2020end}
O.~Mamyrbayev, K.~Alimhan, B.~Zhumazhanov, T.~Turdalykyzy, and F.~Gusmanova,
  ``End-to-end speech recognition in agglutinative languages,'' in
  \emph{Intelligent Information and Database Systems: 12th Asian Conference,
  ACIIDS 2020, Phuket, Thailand, March 23--26, 2020, Proceedings, Part
  II}.\hskip 1em plus 0.5em minus 0.4em\relax Springer, 2020, pp. 391--401.

\bibitem{orken2022study}
M.~Orken, O.~Dina, A.~Keylan, T.~Tolganay, and O.~Mohamed, ``A study of
  transformer-based end-to-end speech recognition system for kazakh language,''
  \emph{Scientific Reports}, vol.~12, no.~1, pp. 1--11, 2022.

\bibitem{beibut2020development}
A.~Beibut, K.~Darkhan, B.~Olimzhan, and K.~Madina, ``Development of automatic
  speech recognition for kazakh language using transfer learning,''
  \emph{International Journal}, vol.~9, no.~4, 2020.

\bibitem{musaev2020use}
M.~Musaev, I.~Khujayorov, and M.~Ochilov, ``The use of neural networks to
  improve the recognition accuracy of explosive and unvoiced phonemes in uzbek
  language,'' in \emph{2020 Information Communication Technologies Conference
  (ICTC)}.\hskip 1em plus 0.5em minus 0.4em\relax IEEE, 2020, pp. 231--234.

\bibitem{ccetinkaya2020improving}
G.~{\c{C}}etinkaya, E.~Ar{\i}soy, and M.~Sara{\c{c}}lar, ``Improving the usage
  of subword-based units for turkish speech recognition,'' in \emph{2020 28th
  Signal Processing and Communications Applications Conference (SIU)}.\hskip
  1em plus 0.5em minus 0.4em\relax IEEE, 2020, pp. 1--4.

\bibitem{ren2022improving}
Z.~Ren, N.~Yolwas, W.~Slamu, R.~Cao, and H.~Wang, ``Improving hybrid
  ctc/attention architecture for agglutinative language speech recognition,''
  \emph{Sensors}, vol.~22, no.~19, p. 7319, 2022.

\bibitem{zeyer2018improved}
A.~Zeyer, K.~Irie, R.~Schl{\"u}ter, and H.~Ney, ``Improved training of
  end-to-end attention models for speech recognition,'' \emph{Proc. Interspeech
  2018}, pp. 7--11, 2018.

\bibitem{graves2006connectionist}
A.~Graves, S.~Fern{\'a}ndez, F.~Gomez, and J.~Schmidhuber, ``Connectionist
  temporal classification: labelling unsegmented sequence data with recurrent
  neural networks,'' in \emph{Proceedings of the 23rd international conference
  on Machine learning}, 2006, pp. 369--376.

\bibitem{li2019improving}
J.~Li, R.~Zhao, H.~Hu, and Y.~Gong, ``Improving rnn transducer modeling for
  end-to-end speech recognition,'' in \emph{2019 IEEE Automatic Speech
  Recognition and Understanding Workshop (ASRU)}.\hskip 1em plus 0.5em minus
  0.4em\relax IEEE, 2019, pp. 114--121.

\bibitem{das2018advancing}
A.~Das, J.~Li, R.~Zhao, and Y.~Gong, ``Advancing connectionist temporal
  classification with attention modeling,'' in \emph{2018 IEEE International
  conference on acoustics, speech and signal processing (ICASSP)}.\hskip 1em
  plus 0.5em minus 0.4em\relax IEEE, 2018, pp. 4769--4773.

\bibitem{kim2017joint}
S.~Kim, T.~Hori, and S.~Watanabe, ``Joint ctc-attention based end-to-end speech
  recognition using multi-task learning,'' in \emph{2017 IEEE international
  conference on acoustics, speech and signal processing (ICASSP)}.\hskip 1em
  plus 0.5em minus 0.4em\relax IEEE, 2017, pp. 4835--4839.

\bibitem{liu2019adversarial}
A.~H. Liu, H.-y. Lee, and L.-s. Lee, ``Adversarial training of end-to-end
  speech recognition using a criticizing language model,'' in \emph{ICASSP
  2019-2019 IEEE International Conference on Acoustics, Speech and Signal
  Processing (ICASSP)}.\hskip 1em plus 0.5em minus 0.4em\relax IEEE, 2019, pp.
  6176--6180.

\bibitem{hori2017joint}
T.~Hori, S.~Watanabe, and J.~R. Hershey, ``Joint ctc/attention decoding for
  end-to-end speech recognition,'' in \emph{Proceedings of the 55th Annual
  Meeting of the Association for Computational Linguistics (Volume 1: Long
  Papers)}, 2017, pp. 518--529.

\bibitem{seki2019vectorized}
H.~Seki, T.~Hori, S.~Watanabe, N.~Moritz, and J.~Le~Roux, ``Vectorized beam
  search for ctc-attention-based speech recognition.'' in \emph{INTERSPEECH},
  2019, pp. 3825--3829.

\bibitem{moritz2019triggered}
N.~Moritz, T.~Hori, and J.~Le~Roux, ``Triggered attention for end-to-end speech
  recognition,'' in \emph{ICASSP 2019-2019 IEEE International Conference on
  Acoustics, Speech and Signal Processing (ICASSP)}.\hskip 1em plus 0.5em minus
  0.4em\relax IEEE, 2019, pp. 5666--5670.

\bibitem{pardede2019generalized}
H.~F. Pardede, V.~Zilvan, D.~Krisnandi, A.~Heryana, and R.~B.~S. Kusumo,
  ``Generalized filter-bank features for robust speech recognition against
  reverberation,'' in \emph{2019 International Conference on Computer, Control,
  Informatics and its Applications (IC3INA)}.\hskip 1em plus 0.5em minus
  0.4em\relax IEEE, 2019, pp. 19--24.

\bibitem{meghanani2021exploration}
A.~Meghanani, C.~Anoop, and A.~Ramakrishnan, ``An exploration of log-mel
  spectrogram and mfcc features for alzheimer's dementia recognition from
  spontaneous speech,'' in \emph{2021 IEEE Spoken Language Technology Workshop
  (SLT)}.\hskip 1em plus 0.5em minus 0.4em\relax IEEE, 2021, pp. 670--677.

\bibitem{ardila2020common}
R.~Ardila, M.~Branson, K.~Davis, M.~Kohler, J.~Meyer, M.~Henretty, R.~Morais,
  L.~Saunders, F.~Tyers, and G.~Weber, ``Common voice: A massively-multilingual
  speech corpus,'' in \emph{Proceedings of the 12th Language Resources and
  Evaluation Conference}, 2020, pp. 4218--4222.

\bibitem{park2019specaugment}
D.~S. Park, W.~Chan, Y.~Zhang, C.-C. Chiu, B.~Zoph, E.~D. Cubuk, and Q.~V. Le,
  ``Specaugment: A simple data augmentation method for automatic speech
  recognition,'' \emph{Proc. Interspeech 2019}, pp. 2613--2617, 2019.

\bibitem{snyder2015musan}
D.~Snyder, G.~Chen, and D.~Povey, ``Musan: A music, speech, and noise corpus,''
  \emph{arXiv preprint arXiv:1510.08484}, 2015.

\end{thebibliography}

\end{document}